\documentclass[aps,superscriptaddress,showpacs,nofootinbib,floatfix,epfs]{revtex4}
\usepackage{amsfonts,amscd,amsmath,amssymb,graphicx,color}
\usepackage[section]{placeins}
\def\rh{r_{\text{\tiny h}}}

\def\erfi{\text{Erfi}}

\def\h0{h_0}
\def\T0{T_{\text{\tiny 0}}}

\def\x0{x_{\scriptscriptstyle 0}}
\def\r0{r_{\scriptscriptstyle 0}}

\def\la0{\lambda_{\scriptscriptstyle 0}}

\def\3q{3\text{\tiny Q}}

\def\4q{4\text{\tiny Q}}
\def\oh{\frac{1}{2}}
\def\ep{\text{e}}
\def\g{\mathfrak{g}}
\def\oh{\frac{1}{2}}

\def\s{\mathfrak{s}}

\def\ci{\text{\tiny (I)}}
\def\cii{\text{\tiny (II)}}
\def\n0{\nu_0}
\begin{document}
\title{More on Polyakov Loops in the Deconfined Phase and Gauge/String Duality}
\author{Oleg Andreev}
 \affiliation{L.D. Landau Institute for Theoretical Physics, Kosygina 2, 119334 Moscow, Russia}
\affiliation{Arnold Sommerfeld Center for Theoretical Physics, LMU-M\"unchen, Theresienstrasse 37, 80333 M\"unchen, Germany}
\begin{abstract} 
We consider a multi-string configuration that provides a new way to compute the expectation value of the Polyakov loop in a five-dimensional framework known as AdS/QCD. The obtained results are in reasonably good agreement with those obtained by lattice simulations for pure $SU(3)$ gauge theory and 
also by the usual single string configuration.
\end{abstract}
\maketitle

\vspace{-7cm}
\begin{flushright}
LMU-ASC 19/17
\end{flushright}
\vspace{6cm}

\section{Introduction} 
\renewcommand{\theequation}{1.\arabic{equation}}
\setcounter{equation}{0}

A pure gauge theory is a theory with only a gauge field. In the case of $SU(3)$, the theory describes the gluonic degrees of freedom of Quantum Chromodynamics (QCD) and, therefore, is a good testbed for work in understanding some perturbative and non-perturbative aspects of the full theory.

There is now a strong evidence base mainly from lattice simulations which shows that in infinite volume, the gluonic matter undergoes a phase transition between the confined and deconfined phases at temperature $T\approx 260\,\text{MeV}$. Universal properties of this transition can be described in terms of the Polyakov loop, which is an order parameter \cite{pol}. Explicitly, it is given by 

\begin{equation}\label{P1}
L_3(T)=\frac{1}{3}\text{tr}P\exp\Bigl[ig\int_0^{1/T}dt\, A_t \Bigr]
\,,
\end{equation}
where the trace is over a fundamental representation $\bf 3$, $t$ is a periodic variable of period $1/T$, with $T$ the temperature, $g$ is a gauge coupling, and $A_t$ is a time component of the gauge field. $P$ denotes path ordering. The usual interpretation of $L_3$ is that it is a phase factor associated to the propagation of a heavy test quark in the fundamental representation of the gauge group, so that the expectation value of the loop is related to the quark free energy.

Special to three colors, there is another fundamental representation which is an anti-triplet, $\bar{\bf 3}$. In fact, it is an anti-symmetric two-index representation of dimension $3$. The relation between the expectation values of the Polyakov loop in these representations is that\footnote{See, for example, \cite{dum}.}

\begin{equation}\label{identity}
\langle\, L_{\bar 3}(T)\,\rangle=\langle\, L^\dagger_3(T)\,\rangle 
\,,
\end{equation}
where $L^\dagger$ is the conjugate of $L$. Since the confined phase is $Z_3$ symmetric, $L_3(T)=0$ below $T_c$. Thus, the above relation is trivially satisfied with $L_{\bar 3}(T)=0$. On the contrary, the global symmetry is broken in the deconfined phase so that $L_3(T)=\ep^{i\phi}L(T)$, where $L(T)=\vert L_3(T)\vert$ and $\ep^{3i\phi}=1$. In this case, the relation yields $L_{\bar 3}(T)=\ep^{-i\phi}L(T)$.

Apart from looking for an ultimate theory that could meet all the challenges of strong coupling, there is a practical way to discuss problems in the physics of strong interactions. It is provided by effective theories. The model we are developing is a simple example of effective string theory. Of course, as any other effective description, it has its own limitations. 

The purpose of the present paper is to further advance the use of effective string theories in QCD. Here we will focus on the Polyakov loop within the five-dimensional effective string model of AdS/QCD. This is a continuation of our previous discussion on pure $SU(3)$ gauge theory \cite{az3,a-pol,a-screen}. In Section II, after reviewing the framework, we present a proposal for how $L_{\bar 3}$ can be incorporated in string models, by involving a baryon vertex  \cite{witten} that is used in higher dimensions as a string-junction \cite{rossi}. Then we compare the results obtained by using this proposal to those of numerical simulations and to our previous results for $L_3$. This serves as a cross check on the model. We close the paper with a few comments in Section III. Some technical issues are treated in the Appendices.

\section{Gauge/string duality and Polyakov loops}
\renewcommand{\theequation}{2.\arabic{equation}}
\setcounter{equation}{0}
\subsection{Preliminaries}

One of the important impacts of the AdS/CFT correspondence \cite{malda} on QCD is that it resumed interest in finding a string description of strong interactions. A five-dimensional\footnote{The reason for the impotence of the fifth dimension was stated in \cite{polyakov}.} effective approach to modeling QCD is known as AdS/QCD \footnote{For a review, see \cite{review} and references therein.}.

We now review the framework in which we will work. Following \cite{az2}, we choose the five-dimensional metric to be of the form

\begin{equation}\label{metric}
ds^2=w(r){\cal R}^2\Bigl(f(r)dt^2+dx^idx^i+f^{-1}(r)dr^2\Bigr)
\,,
\qquad
w(r)=\frac{\ep^{\s r^2}}{r^2}
\,,
\end{equation}
where $f$ is a blackening factor vanishing at $r=\rh$ and $i=1,2,3$. $t$ is a periodic variable of period $1/T$. The boundary of this space, which is $\mathbf{R}^3\times \mathbf{S}^1$, is at $r=0$, where $f=1$. Thus, the geometry in question is a one-parameter deformation, parameterized by $\s$, of the Schwarzschild black hole on $\text{AdS}_5$ space of radius ${\cal R}$. There are good motivations for taking this ansatz, as a version of the soft-wall model of 
AdS/QCD \cite{review}.

Two forms of $f$ were discussed in the context of such a deformation. One, as originally proposed in \cite{az2}, is 

\begin{equation}\label{fS}
f(r)=1-\Bigl(\frac{r}{\rh}\Bigr)^4
\,,
\end{equation}
that is, the blackening factor of the Schwarzschild black hole in $\text{AdS}_5$ space. With this choice, the Hawking temperature is given by 

\begin{equation}\label{TS}
T=\frac{1}{\pi}\sqrt{\frac{\s}{h}}
\,,
\end{equation}
where $h=\s\rh^2$. It is a simple case which enables a great deal of simplification of the resulting equations.

Another one is a solution of a differential equation which relates the functions $w(r)$ and $f(r)$. It follows from the Einstein equations and can be solved explicitly \cite{kir}. As a result, the blackening factor takes the form

\begin{equation}\label{f-kir}
f(r)=1-
\frac{1-\bigl(1+\tfrac{3}{2}\s r^2\bigr)\,\ep^{-\tfrac{3}{2}\s r^2}}
{1-\bigl(1+\tfrac{3}{2}h\bigr)\,\ep^{-\tfrac{3}{2}h}}
\,.
\end{equation}
Given this, the corresponding temperature is 

\begin{equation}\label{T-kir}
T=\frac{9}{8\pi}
\frac{\sqrt{\s}\,h^{\frac{3}{2}}}{\ep^{\tfrac{3}{2}h}-1-\tfrac{3}{2}h}
\,.
\end{equation}
The two blackening factors lead to results which coincide with each other at high temperatures, but display some difference at temperatures close to the critical one.

Apart from the background geometry, another important ingredient for our purposes is the baryon vertex \cite{witten} that describes a coupling of three external charges. In ten dimensions it is a five-brane wrapped over an internal space but in five dimensions it is a point like object (string junction) at which three strings meet. This motives the action in static gauge \cite{3q}

\begin{equation}\label{vertex}
S_{\text{vert}}=\frac{m}{T}\sqrt{f}\,\frac{\ep^{-2\s r^2}}{r}\,,
\end{equation}
where $m$ is a parameter which determines the strength of the gravitational force exerted on the vertex. In fact, what we exploit is one (leading) term from a world-volume action of the brane \cite{3q}. Therefore, $\s$ is nothing else but the deformation parameter defined in \eqref{metric}.

\subsection{String configurations for Polyakov loops}

What are string configurations in five dimensions which correspond to the Polyakov loops? One of those is conceptually simple and well-known. It provides the expectation value of the Polyakov loop in the fundamental (triplet) representation. So what about others? We will in this paper describe a multi-string configuration which provides the expectation value in the anti-triplet representation. In contrast to the first, this one has never been discussed in the literature.

\begin{figure}[htbp]
\centering
\centering
\includegraphics[width=3cm]{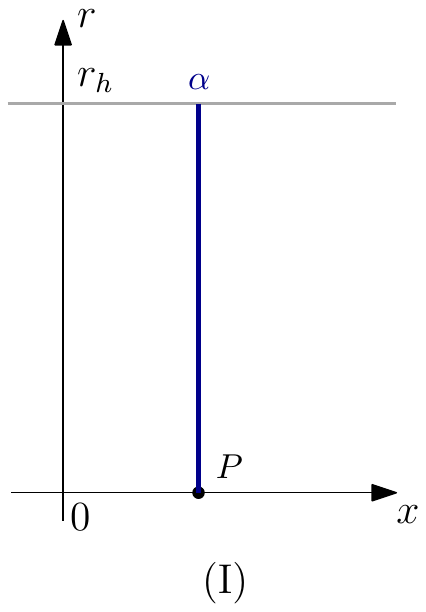}
\hspace{4cm}
\includegraphics[width=3cm]{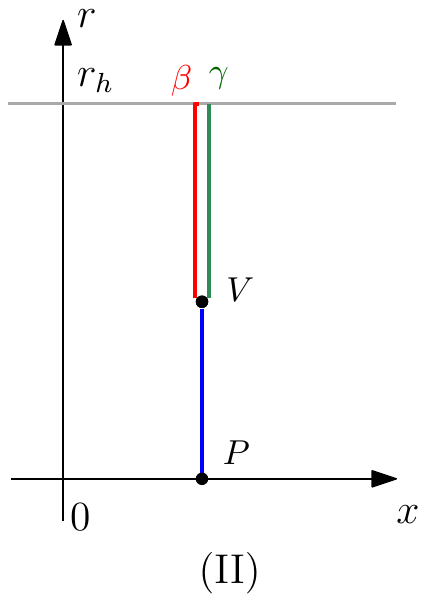}
\caption{{\small Static string configurations. The horizon is at $r=\rh$ and the boundary at $r=0$. The point $P$ represents a quark source and $V$ a baryon vertex. $\alpha$, $\beta$, $\gamma$ denote color indices.}}\label{con}
\label{loops}
\end{figure}

The most straightforward way to see the right configurations is through the factorization property of correlators at infinite loop separation. We begin with the first configuration and therefore consider the correlator of two oppositely oriented loops, $\langle L^\dagger_3 L_3\rangle$.\footnote{Alternatively, one can consider the correlator of two similarly oriented loops, $\langle LL\rangle$ \cite{a-screen}.} A recipe for dealing with a connected string configuration, which is dominant at short distances, was given in \cite{theisen}. Hence one has to put heavy sources (quark and anti-quark) on the boundary of the five-dimensional space, exactly at the positions according to the correlator, and then consider a Nambu-Goto string stretched between these sources. The key point is that gravity pulls the string deeper and deeper into the bulk as the separation between the sources increases. However, when the distance between the sources exceeds a certain critical one, there is no solution that minimizes the Nambu-Goto action \cite{djg}. Roughly speaking, the string breaks as soon as it touches the horizon because the effective string tension vanishes on the horizon \cite{az3}. As a result, the connected configuration transforms into a disconnected one which consists of two single strings. Each single string is similar to that of Figure \ref{loops} on the left. This suggests that the correlator 
can be defined by a sum of contributions from the two string configurations, one connected and one disconnected, so that at distances larger than the critical one only the disconnected configuration 
contributes. Such a definition, however, is not appropriate for pure $SU(3)$ gauge theory as it means that the Debye screening mass is infinite \cite{lattice}. One possible way out seems to be to allow one more connected configuration, like that of \cite{yaffe}, resulting in the desired exponential decay of the correlator at large distances. Importantly, this has no effect on the factorization property of the correlator at infinite separation, where it is completely defined by the disconnected configuration that provides a clear way of computing the expectation value of the Polyakov loop in the fundamental representation.

Next, let us consider the case of three loops, $\langle L_3 L_3 L_3\rangle$. One begins with three heavy sources (quarks) on the boundary of the five-dimensional space, exactly at the loop positions, and then attaches a string to each source. The novelty is the baryon vertex where the strings meet in the bulk. As before, the strings get deeper into the bulk as the relative separation between the sources increases. The main argument is the same as before. To this we only have to answer one question. What happens to the baryon vertex after string breaking? The short and simple answer to the question is that its fate depends on the geometry of the sources. If the quark sources are at the vertices of an equilateral triangle, then on symmetry grounds all the strings break. As a result, the vertex falls into the black hole. In this case the initial configuration simply becomes a disconnected configuration of three single strings. However, if the sources form the symmetric collinear geometry \cite{3q}, then only the side strings break. The middle string remains unbroken and keeps the vertex from falling. Thus, a disconnected configuration consists of two single strings, plus the configuration shown in Figure \ref{loops} on the right. The latter provides another way of computing the expectation value of the Polyakov loop. As before, this is not the whole story as one has to consider additional connected configurations to get a finite screening mass. Those, however, have no effect on the factorization property of the correlator at infinite loop separation. As the configurations of Figure \ref{loops} arise frequently, we denote them by I and II.

How can we interpret both configurations from the group theoretical point of view? This can be done simply by using the analogy between strings and Wilson loops \cite{rossi}. According to that, under $SU(3)$ a string transforms like a Wilson loop, i.e. in the adjoint representation. First consider configuration I. When the string is attached to the quark source, which is in the fundamental representation, one index contracts with the corresponding index of the source, while the second remains uncontracted, as shown in Figure \ref{loops}. It indicates that the source is in the fundamental representation, hence the configuration has to do with the Polyakov loop in the fundamental representation. Of course, what we have done is just suggested how to include color in the five-dimensional description. Further support for this interpretation comes from the explicit calculations made in \cite{a-pol}. Now we go over to configuration II. In doing so, we must bear in mind that in this framework the baryon vertex is associated with the totally antisymmetric tensor. If so, then a little inspection immediately shows antisymmetry under exchange of $\beta$ and $\gamma$, shown in Figure \ref{loops}. Hence it seems natural to interpret this case as that of an anti-symmetric two-index representation, which is the anti-triplet. Further support can be provided by explicit calculations, as we will see in the next subsections.

Given the background geometry, one can calculate the expectation value of a Wilson loop by using the standard approach of AdS/CFT \cite{review}. In practice, it is done by using the saddle point approximation for a string path integral with the boundary conditions that a string world-sheet has the loop for its boundary. The result is written in terms of minimal surfaces (Nambu-Goto actions). It seems natural to apply this approach in our context as well. We do so, by only slightly extending its geometrical interpretation. Thus, the expectation value (absolute) of the Polyakov loop is given by

\begin{equation}\label{bloop}
L(T)=\ep^{-S_{\text{min}}}
\,,
\end{equation}
with $S_{\text{min}}$ the minimal action of a string configuration. For configuration I, $S_{\text{min}}$ is indeed the minimal area of the string world-sheet, but for configuration II it is a sum of three area terms plus a "length" term coming from the baryon vertex. Importantly, the proposal \eqref{bloop} only gives an absolute value. 

\subsubsection{Configuration I}

The single string configuration I is the well-known example in the literature. For the background geometry we are considering, $S_{\text{min}}$ can be calculated analytically, with the result \cite{a-pol}\footnote{See also Appendix A.}

\begin{equation}\label{Pl1}
S^{\ci}=\g\frac{\sqrt{\s}}{T}
\biggl[\sqrt{\pi}\erfi\bigl(\sqrt{h}\,\bigr)
-\frac{\ep^h}{\sqrt{h}}
\,\biggr]+c
\,,
\end{equation}
where $c$ is a normalization constant. $\text{Erfi}(x)$ denotes the imaginary error function. Thus, given the prescription \eqref{bloop}, the expectation value of the Polyakov loop is defined parametrically in terms of $T=T(h)$ and $S^{\ci}=S^{\ci}(h)$. The case $T=\frac{1}{\pi}\sqrt{\frac{\s}{h}}$ is very special, and one can explicitly write down $L(T)$ \cite{a-pol}.

\subsubsection{Configuration II}

We will now describe the multi-string configuration II. In this case, the details are somewhat technical and therefore are given in Appendix A.

The first step is to take one Nambu-Goto string beginning at the heavy source on the boundary and ending on the baryon vertex in the interior and then two others beginning at the vertex and ending on the horizon. Given this, we can write for the total action of the configuration 

\begin{equation}\label{SII}
S^{\cii}=\sum_{i=1}^3 S_i\,+S_{\text{vert}}
\,,
\end{equation}
where $S_i$ is the action of the $i$-string and $S_{\text{vert}}$ is that of the vertex.

The expression for $S_1$ can be read from the formula \eqref{Pc}, while those for the remaining $S_i$ from \eqref{Pa}. Combining all that with the expression for $S_{\text{vert}}$ gives 

\begin{equation}\label{Pl2}
S^{\cii}=\g\frac{\sqrt{\s}}{T}
\biggl[\sqrt{\pi}\Bigl(2 \erfi\bigl(\sqrt{h}\,\bigr)-\erfi\bigl(\nu\sqrt{h}\,\bigr)\Bigr)
+
\frac{1}{\sqrt{h}}\Bigl(\frac{\ep^{h\nu^2}}{\nu}-2\ep{^h}
+
3\kappa\frac{\ep^{-2h\nu^2}}{\nu}\sqrt{f}\,\,\Bigr)
\,\biggr]+c
\,.
\end{equation}
In this derivation we assumed that the vertex is located at $r=\r0$ such that $\nu=\frac{\r0}{\rh}$.

A complete description has to include a gluing recipe which tells us how to glue together the endpoints of strings. In conventional language, this means that a net force exerted on the vertex vanishes. Translating this into mathematical terms gives us the following equation 

\begin{equation}\label{gluing2}
\sqrt{f}+3\kappa\ep^{-3h\nu^2}
\Bigl(1+4h\nu^2-\oh\nu\partial_\nu \Bigr)f
=0
\,,
\end{equation}
where $\kappa=\frac{m}{3\g}$ and $\partial_\nu=\frac{\partial}{\partial\nu}$. To do so we used the results of Appendix A, in particular equation \eqref{gluing}. 

At this point a few comments are in order: (1) We can rewrite \eqref{gluing2} as follows: $\partial_\nu S^{\cii}=0$. The reason is that varying the total action with respect to a vertex location should give a balance of force. Since the problem is in fact one-dimensional and $\r0=\nu\,\rh$, the derivative of $S^{\cii}$ with respect to $\nu$ at fixed $\rh$ determines how $S^{\cii}$ varies with $\r0$. (2) We are unable to solve this equation analytically. In practice it is convenient to take $h$ as a variable and then solve it for $\nu$ numerically. (3) A simple analysis shows that the solution $\nu=\nu(h)$
exists if and only if $\kappa$ takes values in the interval

\begin{equation}\label{kappa}
-\frac{1}{3}<\kappa<0
\,.
\end{equation}
The bounds have a simple geometrical interpretation. At the upper bound, configuration II degenerates into a single string and, hence, becomes configuration I. At the lower bound, it degenerates into two single strings.

Now let us address the issue of stability. Since the configuration is one-dimensional, we will restrict to a small perturbation of $\nu$ (vertex position). In this case, we vary $\nu$ keeping $h$ and $\kappa$ fixed. The first derivative is zero by definition of the equilibrium (gluing condition \eqref{gluing2}). So to explore this question, we have to invoke the second derivative test. 
\begin{figure}[htbp]
\centering
\centering
\includegraphics[width=6.2cm]{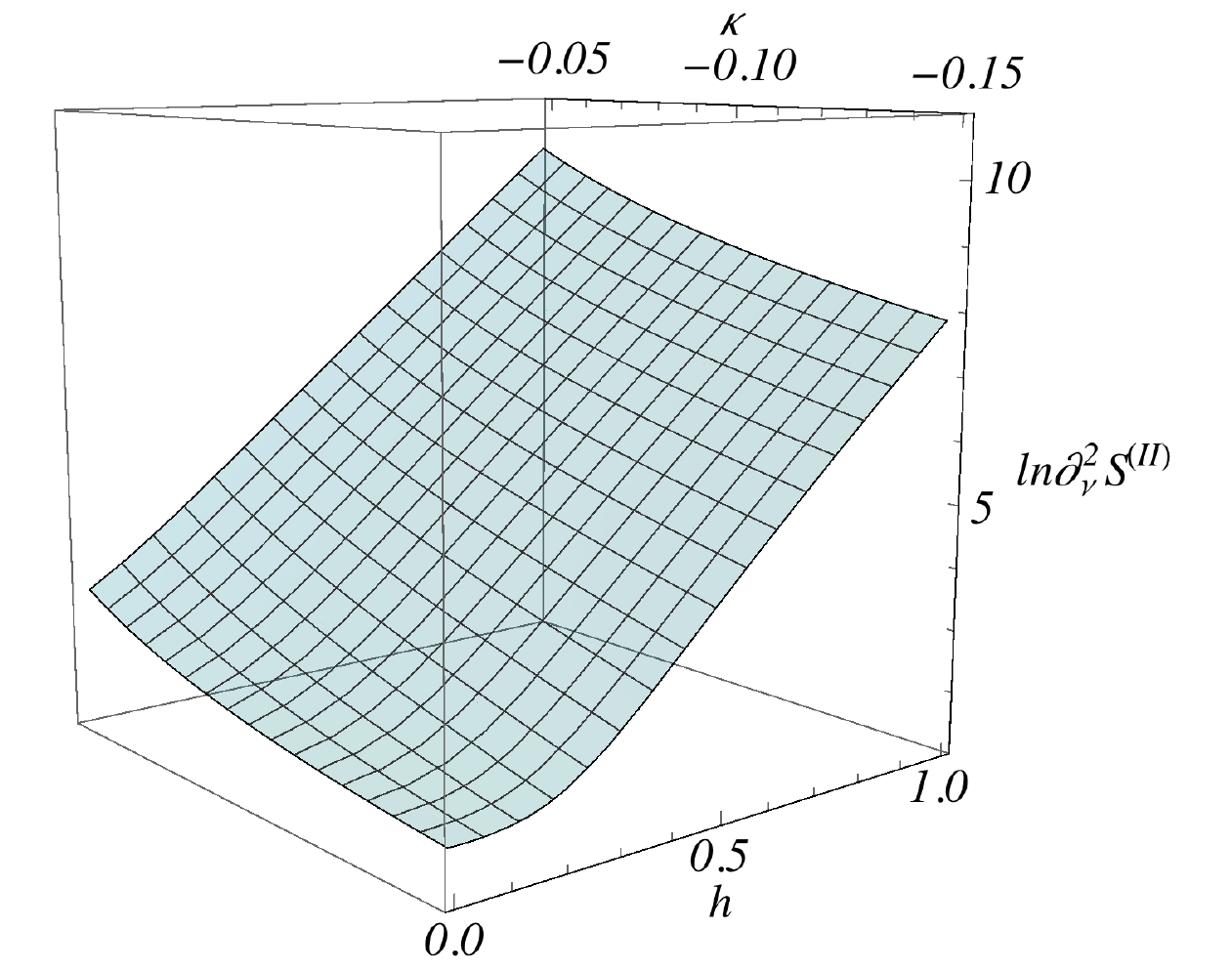}
\hspace{2.25cm}
\includegraphics[width=6.45cm]{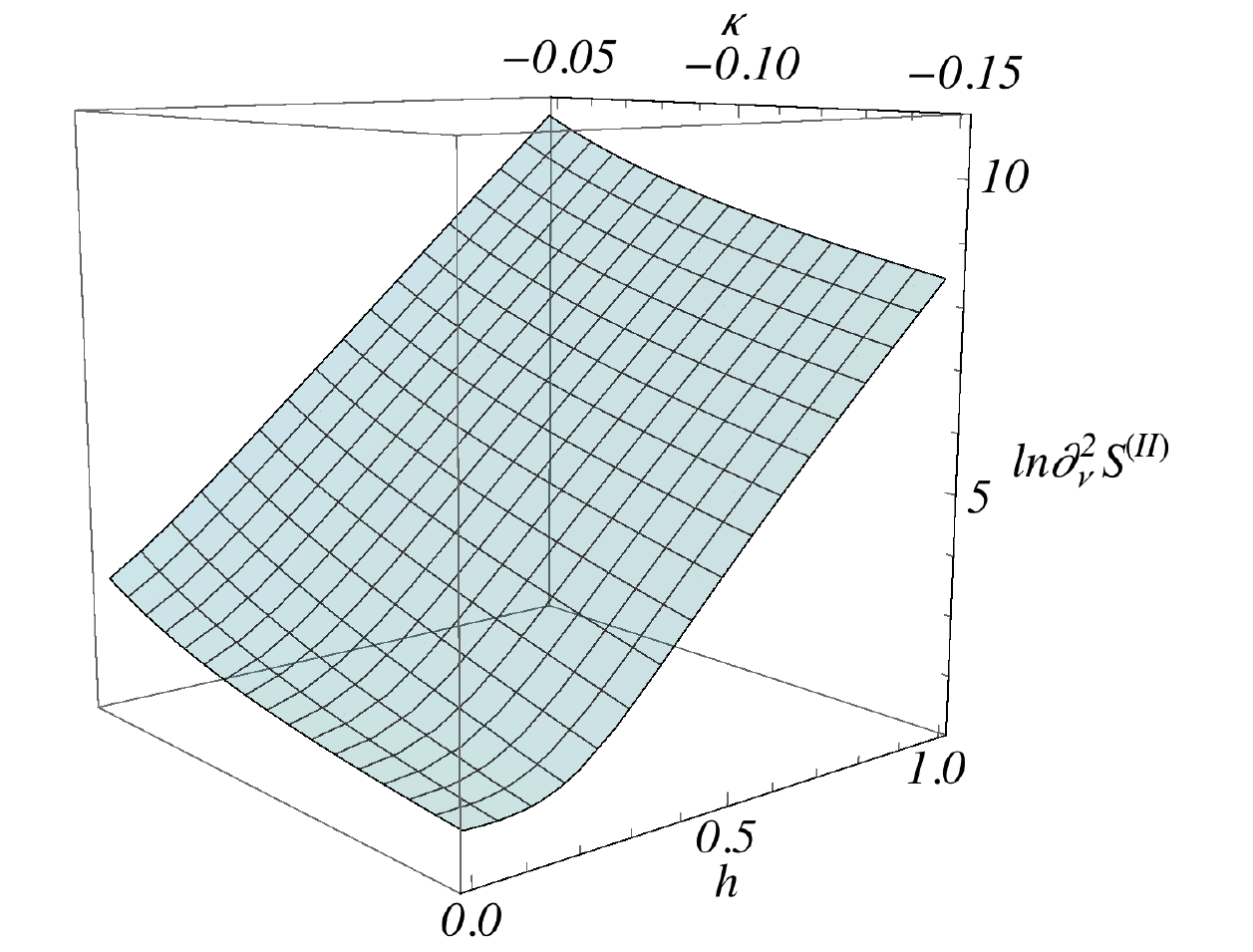}
\caption{{\small $\ln\partial_\nu^2 S^{\cii}$ evaluated on a solution to equation \eqref{gluing2}, for the blackening factor \eqref{fS} on the left and \eqref{f-kir} on the right.}}
\label{stability}
\end{figure}
A little experimentation with Mathematica soon shows that the second derivative evaluated on a solution to equation \eqref{gluing2} is positive. This is 
illustrated in Figure \ref{stability}. Note that, because of the logarithmic scales, it is hard to distinguish one case from another. 

Thus, the expectation value of the Polyakov loop is defined parametrically in terms of $T=T(h)$, $\nu=\nu(h)$, and $S^{\cii}=S^{\cii}(\nu,h)$.

It is noteworthy that in the limit $\s\rightarrow 0$ the expression for $S^{\ci}$ becomes temperature independent, as expected from AdS/CFT. On the other hand, it is not clear what happens with $S^{\cii}$. The problem is that only a single value of the parameter $\kappa$ can be fixed from numerical simulations but this is not enough to determine how it depends on the deformation parameter $\s$.  

\subsection{Numerics}
\begin{figure}[htbp]
\centering
\centering
\includegraphics[width=6.8cm]{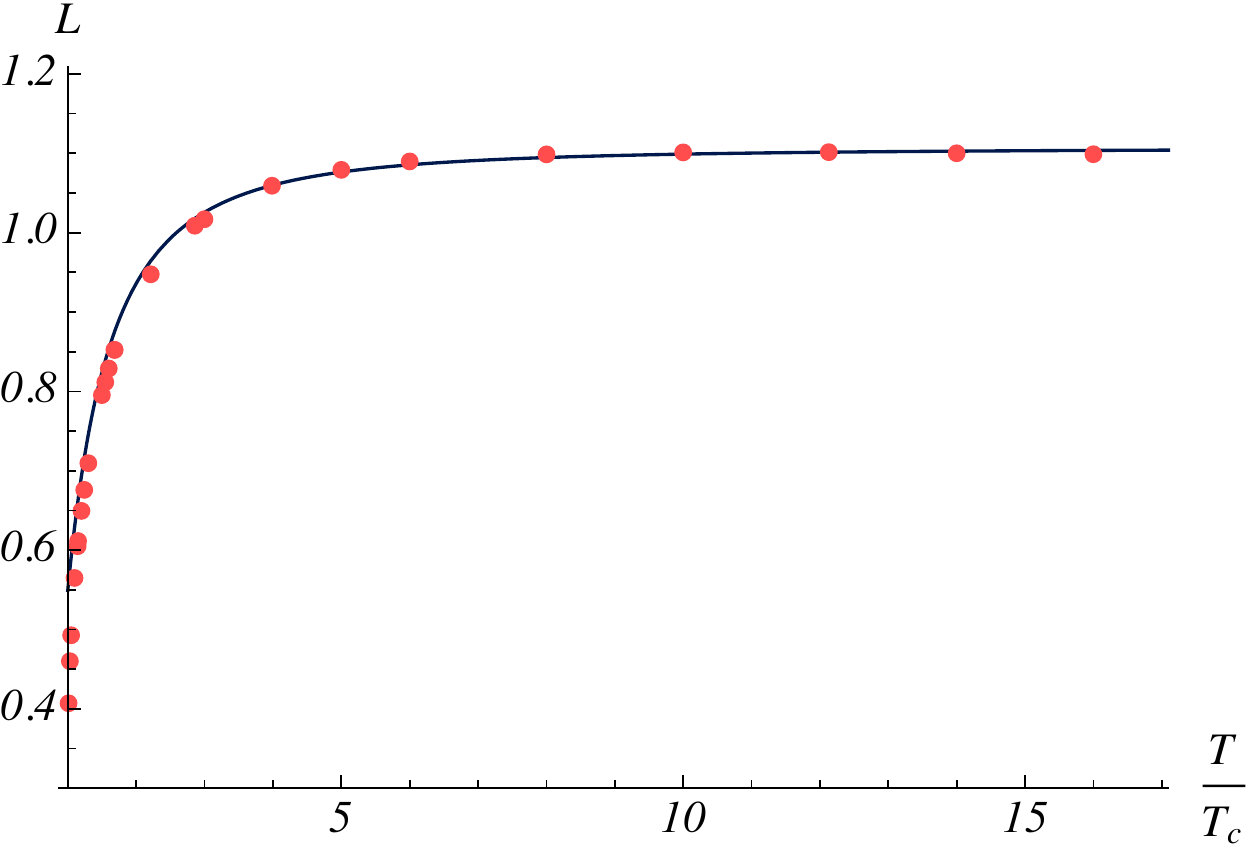}
\hspace{2cm}
\includegraphics[width=6.8cm]{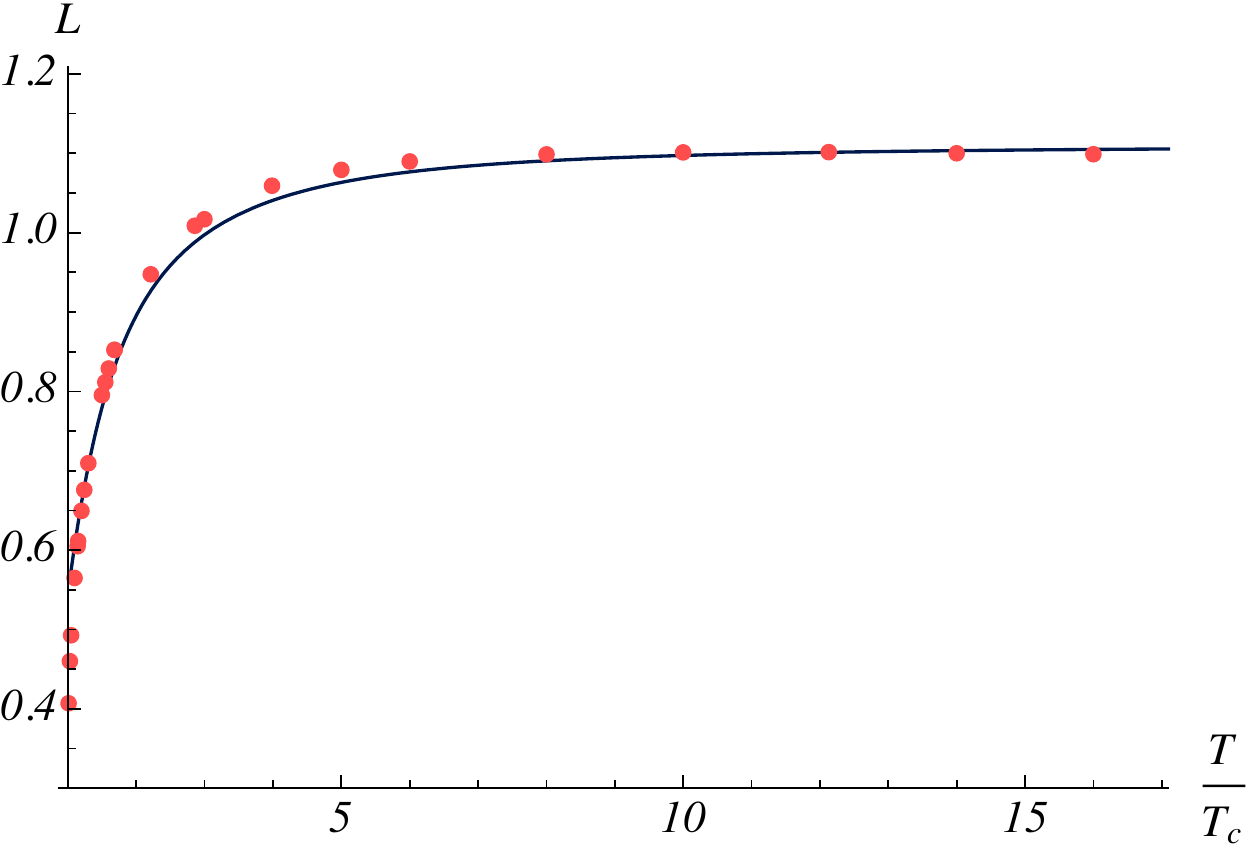}
\caption{{\small The renormalized Polyakov loop in a pure $SU(3)$ gauge theory as a function of temperature. The dots are from lattice simulations of \cite{gupta}. The solid curves show our model results, for the blackening factor \eqref{fS} on the left and \eqref{f-kir} on the right. We do not display any error bars because they are quite small, compatible to the size of the symbols.}}
\label{ploop}
\end{figure}

Having found the parametric expression for the expectation value of the Polyakov loop, we can compare the result with that of numerical simulations. In doing so, we set $\g=0.176$ and $\kappa=-0.083$, i.e., to the {\it same} values as those of \cite{3q} used for modeling the three-quark potential at zero temperature computed on the lattice in the quenched approximation \cite{pdf}. The result is plotted in Figure \ref{ploop}.  
We define $T_c=T\vert_{h=1}$, as follows from the analysis in \cite{az2}, and determine
the value of $c$ from the normalization condition $L=1.1014$ at $T/T_c=12.13$. We see that in general the model reproduces the data quite well, particularly when $f$ is of simple form. This does provide a non-trivial check of the self-consistency of the model we are pursuing.\footnote{Of course, the first thing to check is to make sure that $\kappa=-0.083$ is the interval \eqref{kappa}, but it is obvious.}

\begin{figure}[htbp]
\centering
\centering
\includegraphics[width=6.8cm]{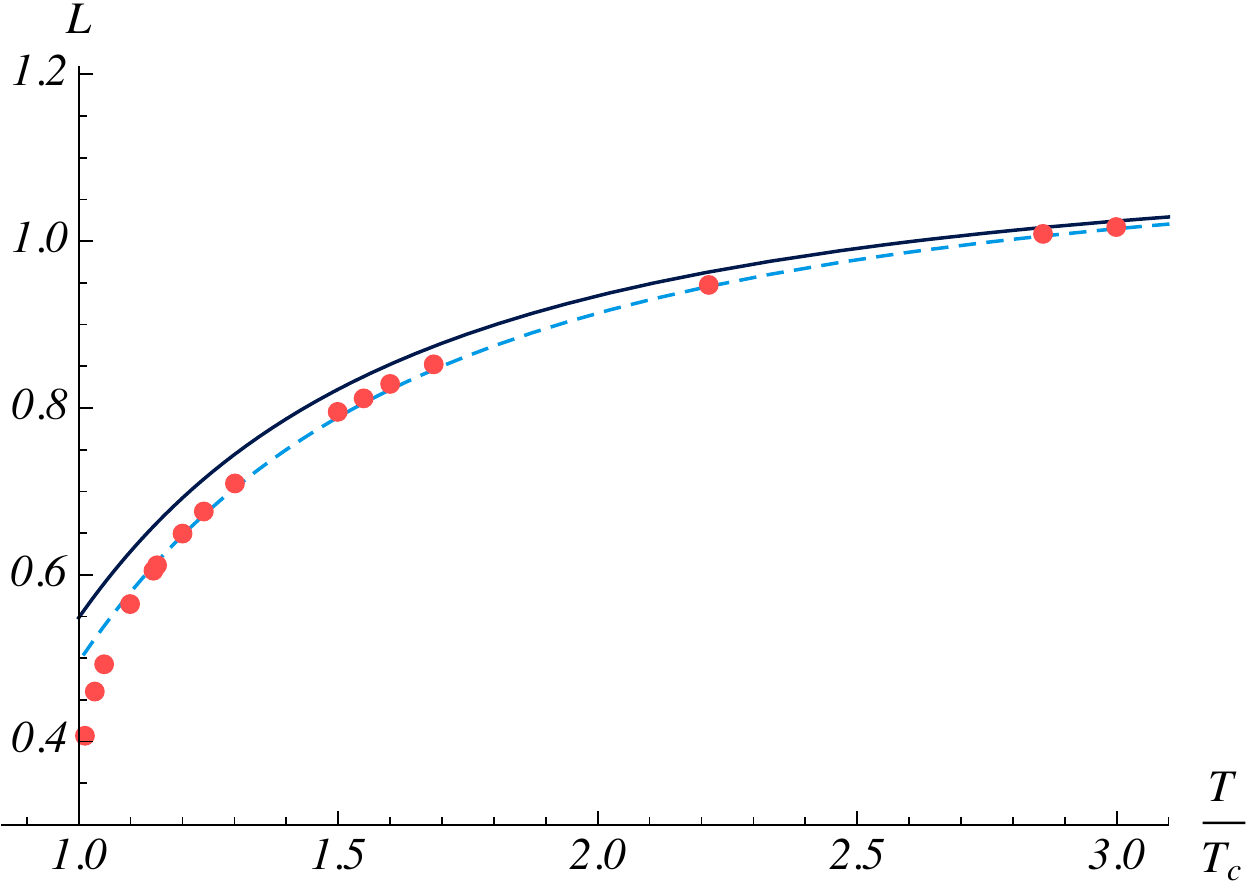}
\hspace{2cm}
\includegraphics[width=6.8cm]{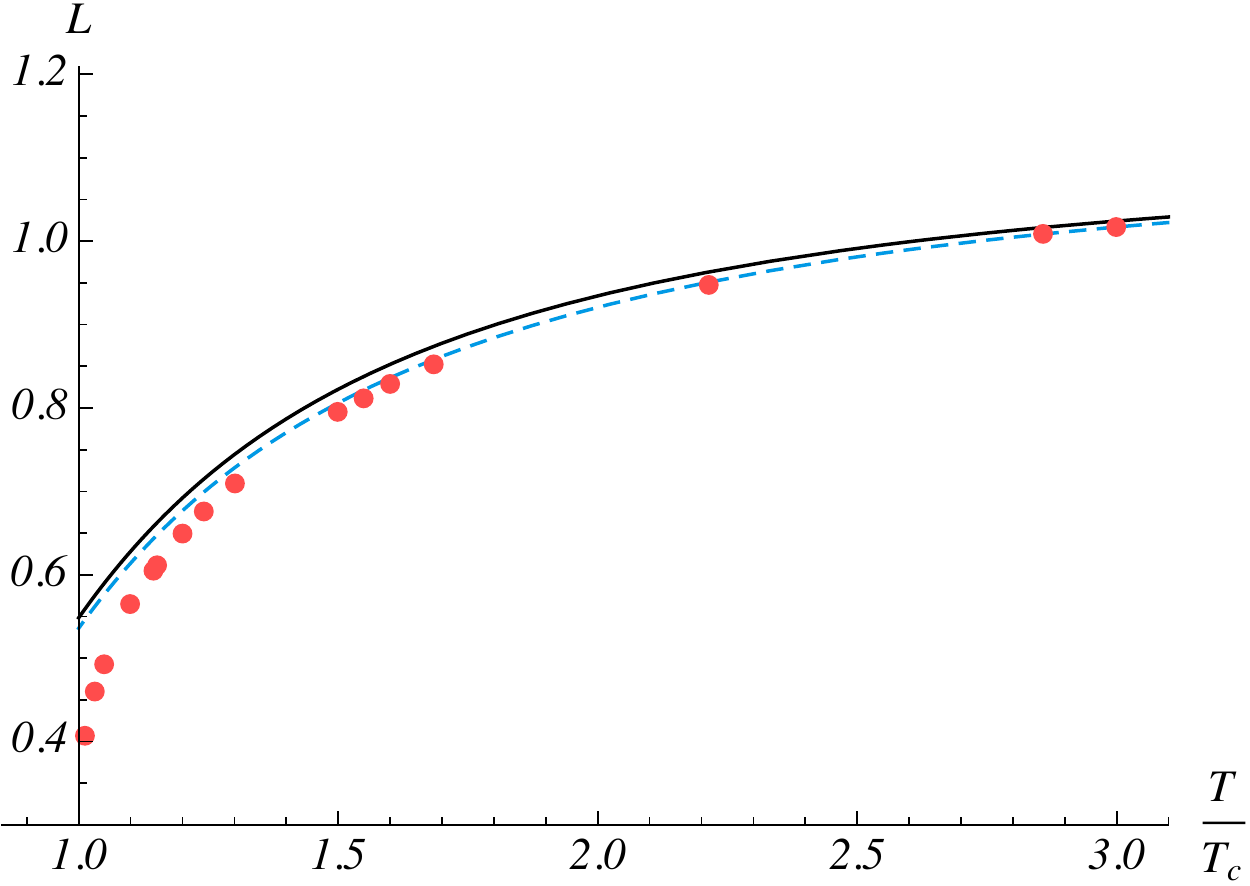}
\caption{{\small The renormalized Polyakov loop in a pure $SU(3)$ pure gauge theory as a function of temperature. In both cases, the dots are from lattice simulations of \cite{gupta}, and the solid curve corresponds to our model at $\g=0.176$ and $\kappa=-0.083$. The dashed curve represents the "best fit" found by letting $\g=0.202$, on the left, and by letting $\kappa=-0.111$ on the right.}}
\label{ploop-fine}
\end{figure}

Actually, the agreement is not so good and becomes worse and worse as the temperature approaches the critical one. For the blackening factor \eqref{fS} this is illustrated in Figure \ref{ploop-fine}. As in the 
case of configuration I \cite{a-pol}, one can improve the fit by changing the value of $\g$, and now also $\kappa$. Doing this helps, except the critical point and its vicinity. One possible explanation is that, since quantum fluctuations become more and more relevant as the temperature tends to the critical one, 
we have to take those into account. We leave untouched this question but hope it will be possible to treat it in the future.

\subsection{More detail on both configurations}

\begin{figure}[htbp]
\centering
\centering
\includegraphics[width=6.8cm]{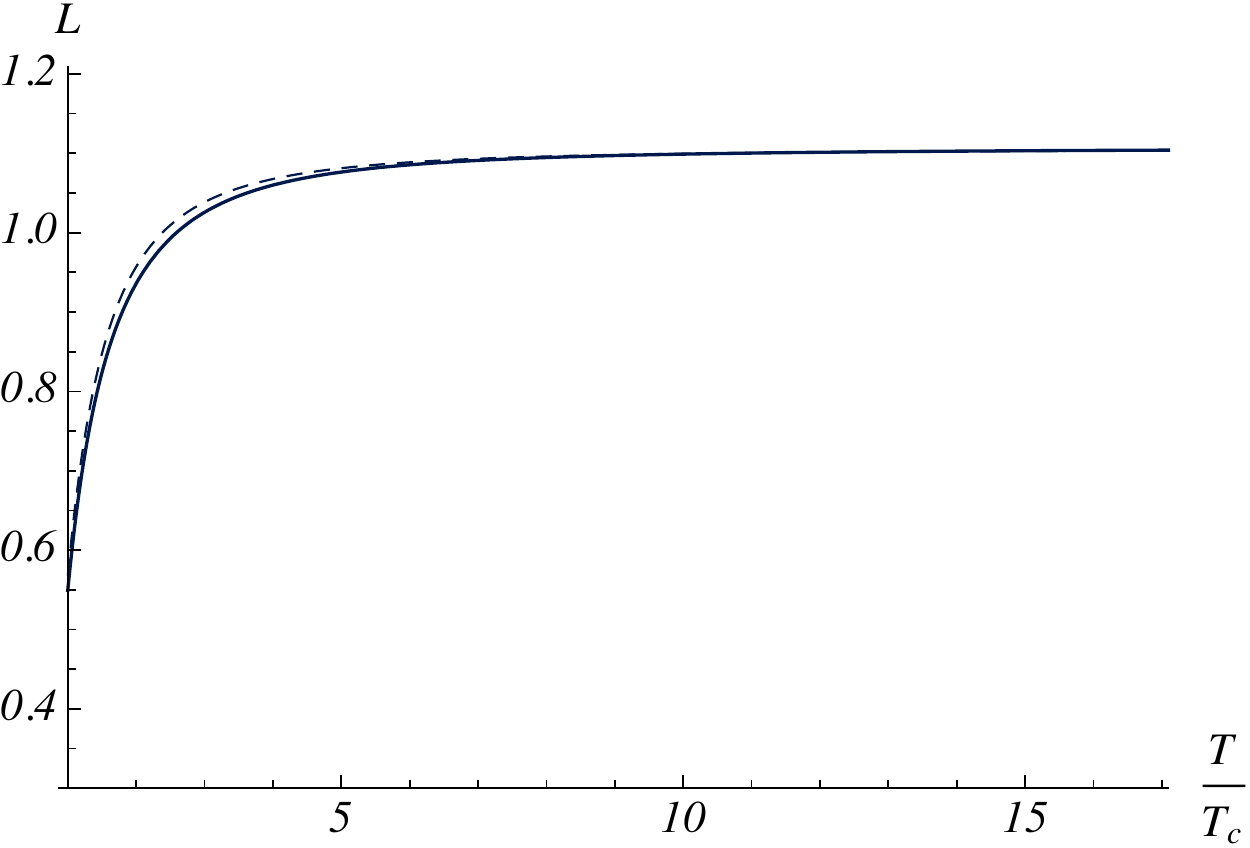}
\hspace{2cm}
\includegraphics[width=6.8cm]{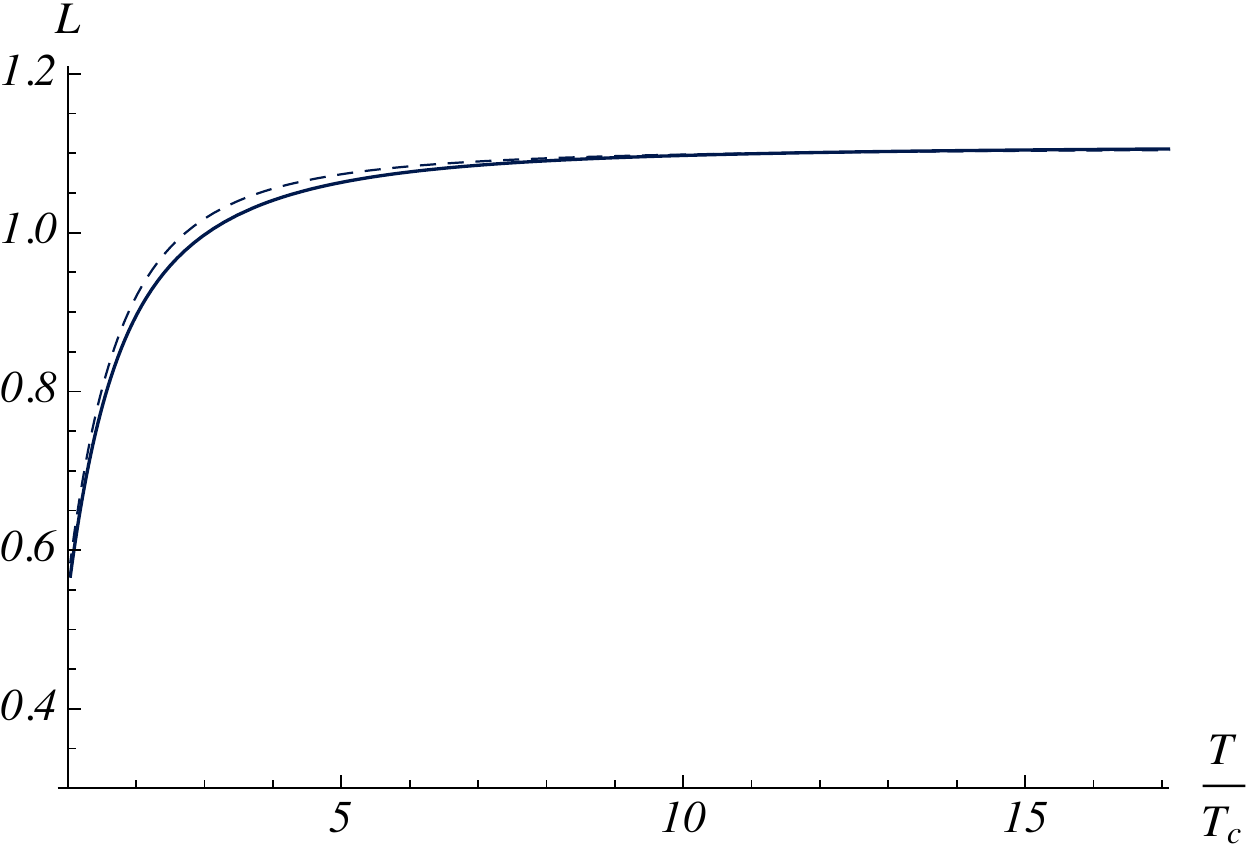}
\caption{{\small $L(T)$ as a function of temperature for the blackening factor \eqref{fS} on the left and \eqref{f-kir} on the right. We set $\g=0.176$ and $\kappa=-0.083$. The dashed curves corresponds to configuration I, while the solid ones to configuration II.}}
\label{ploopI-II}
\end{figure}

It is of primary interest to compare the results obtained here using configuration II with those obtained previously from configuration I \cite{a-pol}. In Figure \ref{ploopI-II} a comparison is shown for both blackening factors. We see that our model is just as   
self-consistent as an effective model can be. In both cases, the only visible discrepancy between the configurations occurs for temperatures in the range of $2T_c$ to $5T_c$. 

In fact, above $2-3\,T_c$ any of our expressions for the expectation value of the Polyakov loop can be satisfactory approximated by the two leading terms in the high-temperature expansion of $S^{(\text{i})}$. Then it takes the form\footnote{Such a form was suggested on phenomenological grounds in \cite{megi}.}

\begin{equation}\label{1/T2}
L(T)\approxeq\exp\biggl[-S_0^{(\text{i})}-S_2^{(\text{i})}\frac{T_c^2}{T^2}\biggr]
\,,
\end{equation}
where $\text{i}=\text{I, II}$. The coefficients $S_0^{(\text{i})}$ and $S_2^{(\text{i})}$ are given by equations \eqref{SfS} and \eqref{Sfkir}, respectively. Let us make a simple estimate of $S_2^{(\text{i})}$ which governs the temperature dependence of $L(T)$ in this range. For $\g=0.176$ and $\kappa=-0.083$, we obtain $S_2^{(\text{I})}\approx 0.553$ and $S_2^{(\text{II})}\approx 0.688$, when $f$ is given by \eqref{fS}, and slightly larger values, $S_2^{(\text{I})}\approx 0.858$ and $S_2^{(\text{II})}\approx 1.269$, when $f$ is given by \eqref{f-kir}. The difference is of order $20-30\,\%$, especially in the latter case. To some extent it is "leveled" in the exponential expression \eqref{1/T2}, because these small coefficients are further suppressed by the factor $T^2_c/T^2$. For lower temperatures the approximation is no longer valid. One has to include the higher order terms in the high-temperature expansion, and this saves the day! The discrepancy between the configurations decreases, as seen from Figure \ref{ploopI-II}.

\section{Concluding Comments}
\renewcommand{\theequation}{3.\arabic{equation}}
\setcounter{equation}{0}

The ability to recover from the five-dimensional string model some quantitative properties of the high temperature gauge theory, in a situation not governed by supersymmetry or conformal invariance, certainly illustrates the power of effective string theories in QCD. Meanwhile, the model rather successfully passes the self consistency test with respect to the previous results on the baryon vertex.

The long-standing question is whether $N_c=3$ is good enough for the $1/N_c$ expansion in QCD. In this paper we consider the tree approximation and do not try to analyze loops. The major reason is that the results of our model being reasonably consistent are in a good agreement with lattice QCD. This indicates that the approximation is good enough and any further refinement of it will be limited to small $1/N_c$ corrections. The secondary reason is technical. The Nambu-Goto formulation is not appropriate for doing so.\footnote{A clue could come from the ten-dimensional Green-Schwarz formulation as proposed for strings on $\text{AdS}_5\times\text{S}^5$ \cite{rr}. However, so far no ten-dimensional string theory dual to QCD has been found.} 

One technical detail seems to be important. The point is that a simple estimate of the critical temperature based on the formula \eqref{TS} gives $T_c=T\vert_{h=1}=\sqrt{\s}/\pi\approx 210\,\text{MeV}$ \cite{az2} that is below $260\,\text{MeV}$ known for a pure $SU(3)$ gauge theory \cite{boyd}. One way out is that the Hawking temperature of the black hole is not exactly equal to the temperature of a dual gauge theory. The two are related by a multiplicative renormalization. According to the estimate of \cite{ssg}, 
such a factor is approximately $3^{\frac{1}{4}}$. With this, one gets $T_c\approx 280\,\text{MeV}$ which is better than before, though still not perfect. We have nothing to say at this point. We were able to avoid this issue because the ratio $T/T_c$ is invariant under rescalings. 

\begin{acknowledgments}
We are grateful to A. Armoni, P.de Forcrand, R. Helling, S. Hofmann, and P. Weisz for helpful discussions and correspondence. We also wish to thank the Arnold Sommerfeld Center for Theoretical Physics and CERN Theory Division for the warm hospitality. This work was supported in part by Russian Science Foundation grant 16-12-10151.
\end{acknowledgments}

\appendix
\section{}
\renewcommand{\theequation}{A.\arabic{equation}}
\setcounter{equation}{0}

The purpose of this Appendix is to provide detailed information, including basic string configurations and a gluing recipe, regarding configurations I and II. Most of the material comes from \cite{a-pol} and \cite{a-screen}. We consider the case when $\rh<\frac{1}{\sqrt{\s}}$, this implies that the corresponding gauge theory is deconfined \cite{az2}.

\subsection{Basic string configurations}

The Nambu-Goto action is given by

\begin{equation}\label{NG}
S=\frac{1}{2\pi\alpha'}\int_0^1 d\sigma\int d\tau\,\sqrt{\gamma}
\,,
\end{equation}
where $\gamma$ is an induced metric on the string world-sheet and $\alpha'$ is a constant which has units of length-squared.

\begin{figure}[htbp]
\centering
\includegraphics[width=3.1cm]{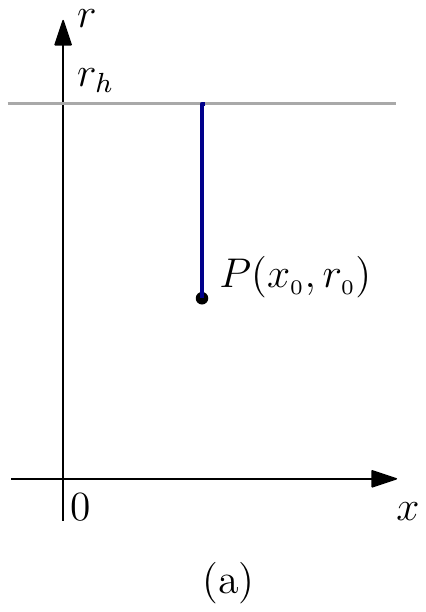}
\hspace{2cm}
\includegraphics[width=3.1cm]{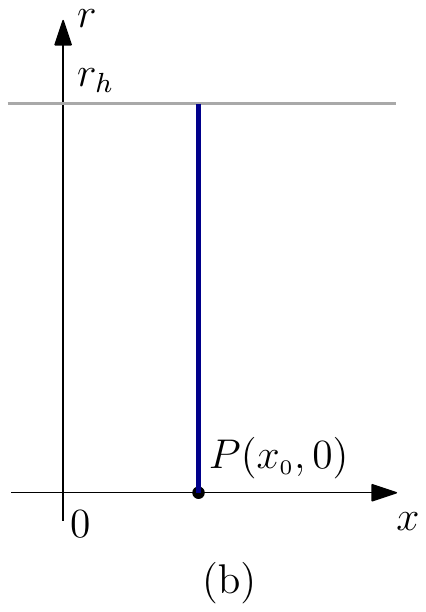}
\hspace{2cm}
\includegraphics[width=3.1cm]{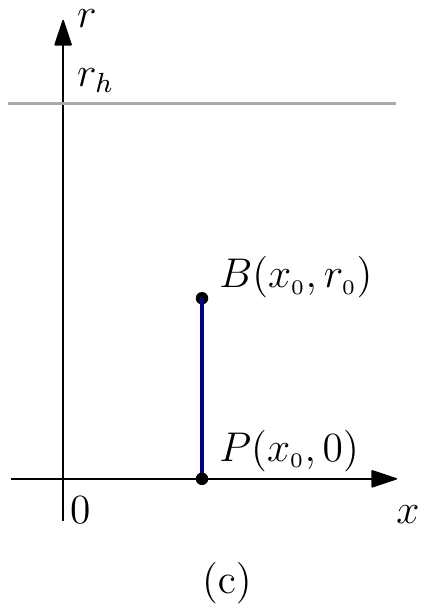}
\caption{{\small Static strings stretched along the $r$-axis.}}
\label{ng}
\end{figure}

Let us consider a string stretched along the $r$-axis, as shown in Figure \ref{ng}. Since we are interested in static configurations, we choose the static gauge $\tau=t$ and $\sigma=r$. For the geometry \eqref{metric}, the action is then 

\begin{equation}\label{NG1}
S=\frac{\g}{T}\int dr\,w\sqrt{1+f(\partial_r x^i)^2} 
\,,
\end{equation}
where $\g=\frac{{\cal R}^2}{2\pi\alpha'}$. Using this action, one of the solutions of the equation of motion for $x^i$ is simply $x^i=const$ \cite{a-pol} that represents a straight string stretched along the $r$-axis, say between the points $r_1$ and $r_2$. 
Given such a solution, we can easily evaluate $S$ for it, with the result 

\begin{equation}\label{master}
S=\frac{\g}{T}\int_{r_1}^{r_2} dr\,w(r)=
\frac{\g}{T}
\biggl[\sqrt{\pi\s}\Bigl(\erfi\bigl(\sqrt{\s}r_2\bigr)-\erfi\bigl(\sqrt{\s}r_1\bigr)\Bigr)
-\frac{\ep^{\s r_2^2}}{r_2}
+\frac{\ep^{\s r_1^2}}{r_1}
\,
\biggr]
\,.
\end{equation}
Here we have used the explicit form of the warp factor. 

Let us now apply this formula to the configurations of Figure \ref{ng}. If $r_1=\r0\not=0$ and $r_2=\rh$, as in Figure \ref{ng}(a), then the action takes the form

\begin{equation}\label{Pa}
S^{(\text{a})}=\g\frac{\sqrt{\s}}{T}
\biggl[\sqrt{\pi}\Bigl(\erfi\bigl(\sqrt{h}\,\bigr)-\erfi\bigl(\nu\sqrt{h}\,\bigr)\Bigr)+
\frac{1}{\sqrt{h}}
\Bigl(\frac{\ep^{h\nu^2}}{\nu}
-\ep^{h}\Bigr)
\biggr]
\,,
\end{equation}
with $h=\s\rh^2$ and $\nu=\frac{\r0}{\rh}$.

For $r_1=\r0=0$, as in Figure \ref{ng}(b), the last term in \eqref{master} is singular and it requires regularization. We do so by imposing a cutoff $\epsilon$ on the low limit of 
integration. Then subtracting the $\frac{1}{\epsilon}$ term and letting $\epsilon=0$, we get a renormalized action \cite{a-pol}
 
\begin{equation}\label{Pb}
S^{(\text{b})}=\g\frac{\sqrt{\s}}{T}
\biggl[\sqrt{\pi}\erfi\bigl(\sqrt{h}\,\bigr)
-\frac{\ep^h}{\sqrt{h}}
\,\biggr]+c
\,,
\end{equation}
where $c$ is a normalization constant. What it describes is configuration I.

Once $S^{(\text{a})}$ and $S^{(\text{b})}$ are given, we can find the remaining $S^{(\text{c})}$ by simply taking the difference of those, as readily seen from Figure \ref{ng}(c). This leads to

\begin{equation}\label{Pc}
S^{(\text{c})}=\g\frac{\sqrt{\s}}{T}
\biggl[-\sqrt{\pi}\erfi\bigl(\nu\sqrt{h}\,\bigr)+
\frac{\ep^{h\nu^2}}{\nu\sqrt{h}}
\biggr]+c
\,.
\end{equation}
\subsection{Gluing condition}

Having the basic configurations of Figure \ref{ng} is necessary but it is not sufficient for our purposes. What we also need to know is a gluing recipe which describes how configuration II of Figure \ref{loops} can be obtained by gluing together three strings joining at a baryon vertex. This is easy to understand within a five-dimensional framework, where a baryon vertex looks like a point-like object. Here is a natural recipe. Any static configuration must obey the condition that a net force vanishes at each vertex. So\footnote{We denote five-dimensional vectors by boldface letters.}

\begin{equation}\label{v-netforce}
\mathbf{f}+\sum_{i=1}^3\mathbf{e}_{i}=\mathbf{0}
\,,
\end{equation}
where $\mathbf{f}$ is a gravitational force and $\mathbf{e}_i$ is a force exerted by the $i$ string on the vertex. This formula can be obtained by varying the action, which is a sum of three Nambu-Goto actions and a vertex action, with respect to a location of the vertex. 

\begin{figure}[htbp]
\includegraphics[width=3.4cm]{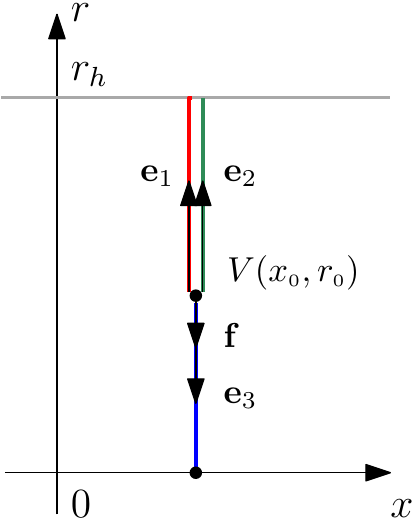}
\caption{{\small The balance of force at $V$. The gravitational force is directed towards the boundary.}}
\label{netforce}
\end{figure}

For present purposes, we are interested in the balance of force at the vertex of configuration II, as sketched in Figure \ref{netforce}. Since all the strings are stretched along the $r$-axis, the equations 
for the $t$ and $x^i$-components of the net force are trivially satisfied. Thus, we need only to consider that for the $r$-component. Following \cite{a-screen}, it is convenient to normalize the vectors $\mathbf{e}_{i}$ such that the gravitational force has the $r$-component $\mathbf{f}^r=-3\kappa\nu^2\ep^{-h\nu^2}\partial_{\nu}\bigl(\frac{\ep^{-2h\nu^2}}{\nu}\sqrt{f}\bigr)$, with $\kappa=\frac{m}{3\g}$ and $\partial_\nu=\frac{\partial}{\partial\nu}$. Finally, the equation for the $r$-component can be written in the form

\begin{equation}\label{gluing}
1-3\kappa\nu^2\ep^{-h\nu^2}\partial_{\nu}
\Bigl(\frac{\ep^{-2h\nu^2}}{\nu}\sqrt{f}\Bigr)=0
\,.
\end{equation}
An important point is that gravity pulls the vertex toward the boundary, otherwise this equation has no solutions and, as a consequence, the configuration is unstable.

\section{High-temperature expansion}
\renewcommand{\theequation}{B.\arabic{equation}}
\setcounter{equation}{0}

This appendix collects together some of the formulas which are used in analyzing the high-temperature expansion of $S^{\ci}$ and $S^{\cii}$ in Section II.

First, we need a simple fact about the parametric equations \eqref{fS} and \eqref{f-kir}: high $T$ corresponds to small $h$. In the geometric context, this means that the horizon approaches the boundary as the time interval shrinks to zero. It is manifest for \eqref{fS}, and requires not much more effort for \eqref{f-kir}.

Second, $S^{\ci}$ and $S^{\cii}$ are the functions of $h$, and each of those can be expanded in a Taylor series about $h=0$. In the case of $S^{\ci}$, it can be shown using equations \eqref{fS} and \eqref{f-kir}. The case of $S^{\cii}$ is a bit more involved, and proceeds by seeking a solution of \eqref{gluing} in the form of a power series about $h=0$, $\nu=\sum_{n=0}^\infty \nu_n h^n$. Combining it with equations \eqref{fS} and \eqref{f-kir} leads to the desired result.

Finally, expressing $h$ in terms of $T$, $h=\frac{\s}{\pi^2T^2}\sum_{n=0}^\infty h_n T^{-2n}$, and putting all together, we find that the high-temperature expansion of $S$, as a power series, involves only non-positive powers of $T^2$. So for high enough temperature, 

\begin{equation}\label{ST}
S^{({\text i})}(T)=\sum_{n=0}^\infty S_{2n}^{({\text i})}\biggl(\frac{T_c}{T}\biggr)^{2n}
\,,
\end{equation}
where ${\text i}=\text{I, II}$. As before, we define $T_c=T\vert_{h=1}$. It is straightforward to compute the coefficients in this series explicitly. For the leading coefficients, the results are given below.

We start with the blackening factor \eqref{fS}. In this case, we get  

\begin{equation}\label{SfS}
S^{\ci}_0=c-\pi\g
\,,
\qquad
S^{\ci}_2=\pi\g
\,,
\qquad
S^{\cii}_0=c-2\pi\g\Bigl(1-\frac{\n0^3}{1+\n0^4}\Bigr)
\,,
\qquad
S^{\cii}_2=\pi\g\Bigl(2+\n0\frac{1-3\n0^4}{1+\n0^4}\Bigr)
\,,
\end{equation}
where $\n0=\Bigl(\frac{4}{1+\sqrt{1+72\kappa^2}}-1\Bigr)^{\frac{1}{4}}$.

For the blackening factor \eqref{f-kir}, the corresponding coefficients are given by 

\begin{equation}\label{Sfkir}
S^{\ci}_0=c-\pi\g
\,,
\qquad
S^{\ci}_2=\gamma_0\pi\g
\,,
\qquad
S^{\cii}_0=c-2\pi\g\Bigl(1-\frac{\n0^3}{1+\n0^4}\Bigr)
\,,
\qquad
S^{\cii}_2=\gamma_0\pi\g\Bigl(2+\n0\frac{2+3\n0^2-7\n0^4}{1+\n0^4}\Bigr)
\,,
\end{equation}
with $\gamma_0=\frac{32}{81}\Bigl(\ep^{\frac{3}{2}}-\tfrac{5}{2}\Bigr)^2$. Note that in both cases the $S_0^{(\text i)}$'s are the same, since the blackening factors coincide in the $\s\rightarrow 0$ limit.


\end{document}